# Intelligent Operation and Maintenance and Prediction Model Optimization for Improving Wind Power Generation Efficiency

Xun Liu
*Faculty of Education*
*Shinawatra University*
Bangkok, Thailand
lxhkj2011@gmail.com

Xiaobin Wu
*School of AutomotiveEngineering*
*Chengdu Industry And Trade College*
Chengdu, China
wuxiaobin@cdgmxy.edu.cn

Jiaqi He
*Tilburg School of Social and Behavioral Sciences*
*Tilburg university*
Tilburg, The Netherlands
hjqhejiaqi@gmail.com

Rajan Das Gupta
*Department of Computer Science and Engineering*
*Jahangirnagar University*
Dhaka, Bangladesh
rajandasgupta.me@gmail.com

*Abstract*—This study explores the effectiveness of predictive maintenance models and the optimization of intelligent Operation and Maintenance (O&M) systems in improving wind power generation efficiency. Through qualitative research, structured interviews were conducted with five wind farm engineers and maintenance managers, each with extensive experience in turbine operations. Using thematic analysis, the study revealed that while predictive maintenance models effectively reduce downtime by identifying major faults, they struggle with detecting smaller, gradual failures. Key challenges, such as false positives, sensor malfunctions, and the integration of new models with older turbines, were also identified. Advanced technologies like digital twins, SCADA systems, and condition monitoring have enhanced turbine maintenance but require further improvements, particularly in AI refinement and real-time data integration. The findings highlight the need for continued development to optimize wind turbine performance and support the broader adoption of renewable energy.

## I. INTRODUCTION

Wind energy is one of the most important sources of renewable energy used to mitigate the climate change effects; however, increasing the efficiency of the wind turbines is still an engineering issue. It has been found that predictive maintenance where machine learning and data analytics are used has the potential of making huge difference in the reliability and performance of turbines. For example, Schlechtingen and Santos [1] showed that it is possible to monitor the condition of the system by the means of Supervisory Control and Data Acquisition (SCADA) data to predict failures and minimise the time of breakdowns. Such models enable early fault detection, reduce maintenance expenses and improve energy generation. Furthermore, there is an emergence of intelligent operation and maintenance (O&M) systems to monitor and optimize turbines in operation [2]. Therefore, for the wind energy sector, as the market expands further more sophisticated and efficient predictive models and smart O&M strategies are crucial for achieving better results and cutting further costs down.

### A. Problem Statement

The efficiency of wind power generation is often hindered by the high operational and maintenance costs associated with unexpected turbine failures and system downtime. Traditional maintenance strategies are reactive, leading to costly repairs and prolonged downtime, which negatively affect energy output. Moreover, the current predictive models used in wind farms are often not optimized for real-world conditions, limiting their effectiveness. There is a critical need to enhance predictive maintenance and optimize intelligent operation and maintenance (O&M) systems to ensure continuous, efficient wind power generation while minimizing operational disruptions. Hence, in this context the objectives of this study are given below:

### B. Study Objectives

- To analyze the effectiveness of predictive maintenance models in reducing downtime and improving the operational efficiency of wind turbines.
- To explore optimization strategies for intelligent operation and maintenance (O&M) systems to enhance wind power generation efficiency

### C. Significance of the Study

The significance of this study lies in the fact that it can help improve the predictive maintenance models as well as the intelligent O&M systems essential in reducing turbine downtime and costs of operating wind power generators. Thus, improving these systems, the study helps to increase the efficiency and stability of wind energy generation, therefore, promoting sustainable energy development worldwide.

## II. LITERATURE REVIEW

### A. Predictive Maintenance Models for Reducing Downtime in Wind Turbines

The application of predictive maintenance (PdM) models in wind turbines is critical for reducing system downtime by predicting component failures before they occur. These models typically rely on various forms of data, including vibration, temperature, and operational data, to monitor the health of the turbine and its components. By leveraging advanced algorithms, PdM systems can analyze this data to detect early signs of wear and tear, allowing maintenance to be performed proactively rather than reactively.

For example, Supervisory Control and Data Acquisition (SCADA) systems, which collect real-time operational data,

have been widely used to monitor wind turbine conditions and predict failures. Zaher et al[3]. (2009) demonstrated how SCADA data could be analyzed to predict potential failures, allowing maintenance teams to schedule repairs before a breakdown occurs. This approach reduces downtime and maintenance costs while improving energy generation efficiency. However, predictive models based solely on SCADA data face significant challenges, such as noisy sensor readings and missing data, which can compromise their reliability.

To address these issues, researchers have increasingly explored the integration of advanced techniques, such as Artificial Neural Networks (ANN), deep learning, and ensemble learning, to enhance the accuracy of PdM models. For instance, Wuest et al[4]. (2016) proposed an ensemble learning-based approach that combines multiple machine learning algorithms to improve fault detection accuracy. This method has shown promising results in identifying major component failures, such as rotor blade faults, with greater precision and speed. However, these models require continuous updates to adapt to the changing conditions of turbines and their operational environments, and they often come with high computational costs.

The integration of data-driven and knowledge-driven techniques is increasingly recognized as a critical approach to improving the reliability of PdM architectures and reducing false positive alarms. Cheng et al. (2022) emphasize that combining these techniques allows for more accurate and robust fault detection. Moreover, the use of visual analytics, particularly when displayed via intuitive dashboards, enhances the ability of maintenance personnel to respond promptly and identify effective preventive measures. These dashboards provide real-time insights into turbine health, enabling maintenance teams to prioritize actions based on the severity of detected anomalies[17].

Despite the advancements in PdM algorithms over the past decades, a significant challenge persists: the lack of labeled failure data in the manufacturing industry. This limitation restricts the development and validation of predictive models, ultimately affecting their real-world applicability. Addressing these challenges is crucial for ensuring the seamless implementation of PdM with minimal human intervention (Cheng et al., 2022)[17]. As such, more research is needed to improve data quality, reduce false alarms, and enhance the integration of real-time data with predictive maintenance models.

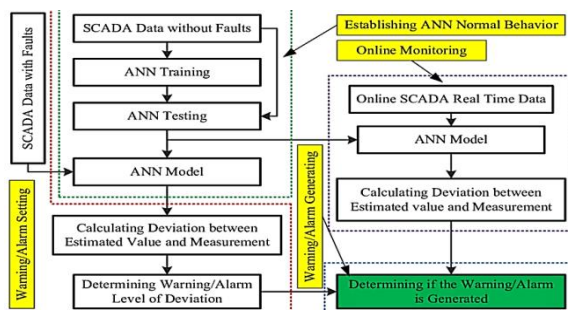

Fig. 1. Procedure of fault prediction based on SCADA and ANN (Source: https://www.researchgate.net/figure/Procedure-of-fault-prediction-based-on-SCADA-and-ANN_fig3_326013496)

For instance, Wuest et al. [4] developed an ensemble learning-based technique for accurate fault detection in the turbine based on multiple machine learning algorithms. This strategy proved much more effective compared to the conventional methods in analyzing failures of major components, like rotor blades, with higher precision and in less time. However, there are still issues with such approaches to real wind farms, mainly, related to the computational costs and the fact that these models have to be updated through time to adapt to the changing conditions of the turbines.

### B. Optimization of Intelligent O&M Systems for Wind Power Efficiency

Optimizing intelligent O&M systems is another key factor that defines the improvement of wind turbines by means of data analysis and assessment of failures. In conventional methods, operation and maintenance are mostly detective in nature focusing on turbine problems after they develop into a failure causing expensive time-outs. While, the intelligent operation and maintenance systems use the artificial neural network and big data analysis for failure prediction and best probable time of maintenance. For instance, Feng et al [6] analyzed how machine-learning methods could be used for improving maintenance schedule and decrease the time that the turbines are out of service while increasing energy generation.

The advanced tools for enhancing operation and maintenance systems are the digital twin technology that produces the realistic models of the real turbines to perform under the real operating conditions. Digital twins help operators detect faults in advance and make decisions identifying potential for improving utilisation. Tao et al [7] pointed out that when digital twin was incorporated with AI algorithms, it increased efficiency in fault detection, lowered maintenance expenses, increased energy yield through tweaking the operation of the turbine when it was in line with differing environmental conditions.

Nevertheless, some challenges are still present in implementing intelligent operations and maintenance systems. According to Zhang et al. [8] the deployment of Advanced AI-enabled systems on a greater scale, substantial volumes of high-quality data as well as high processing power are necessary and make it expensive. Furthermore, the incorporation of real time data from various sources, including weather conditions and turbine health also poses another challenge in developing complex model, which cannot be scalable. However, these challenges can be addressed by the sustained evolution of hybrid techniques that integrate real-time monitoring data and the historical record to enhance decision-making on wind turbine reliability and efficiency.

### C. Literature Gap

Despite the positive results shown by the predictive maintenance models and intelligent O&M systems, some issues would still be faced when applying these solutions in large turbines. Current literature has ignored the need to fine-tune these models for different types of turbines and various operating environments. Furthermore, methods for real-time and big data integration spanning across different application domains that are cost-effective and easily scalable are yet to be developed fully. Therefore, the purpose of this research is to address these gaps by investigating the potential of predictive maintenance for improving O&M systems' efficiency in wind power generation and identifying possible improvements for intelligent O&M systems.

### III. METHODOLOGY

In this study, the research method adopted is qualitative and involves the analysis of predictive maintenance models and O&M optimisation of wind power systems. Data was obtained through the use of structured interviews with key informants comprising of three wind farm engineers and two maintenance managers. The thematic analysis was performed in an attempt to find out the patterns and the insights within the gathered data. This approach offers a comprehensive picture of practice-based reality. Fig. 2 illustrates the visual representation of the research methodology proposed for the current study.

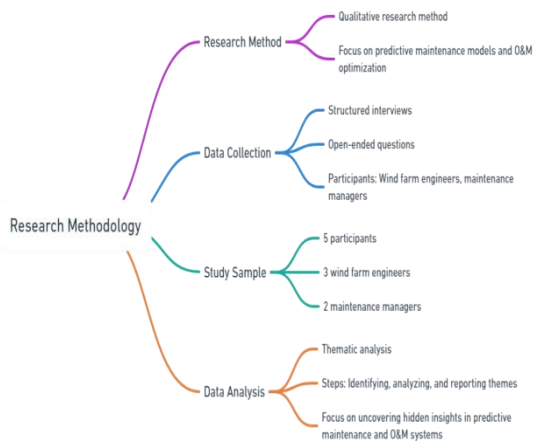

Fig. 2. Methodology overview (Source: Authors)

#### A. Research Method

In this research, a qualitative research approach was used in order to identify the key predictive maintenance models and optimization techniques needed in intelligent O&M systems to improve the efficiency of wind power generation. The rationale for employing the qualitative method lies in the assumption that this research approach enables the establishment of a profound understanding of the participants' experiences and their visions. This method gives detailed information to context that make it easier to discover other difficulties that may not be revealed through quantitative techniques. As a result, qualitative research is the most suitable choice as it seeks to understand the specifics of the systems in use as well as the subjects' individual perceptions

#### B. Data Collection

Data for this research was therefore collected using structured interviews with selected professionals in the wind power operations and maintenance sector. The interviews took general form and focused on general questions, where respondents could give a lot of details on Predictive maintenance models and O&M optimization. It helps the researcher to collect rich data and views in relation to the topic and therefore enhance the understanding of the problem at hand. The participants were questioned about problems, efficiency and further developments in predictive maintenance and operations & maintenance systems.

The primary data was collected from the wind farm engineers and maintenance managers because they are proactive in the monitoring of the turbine efficiency as well as in the formulation of the maintenance techniques to be adopted. These professionals can give a more pragmatic understanding of the operational use of intelligent O&M systems, as well as discussing the shortcomings of the typical predictive maintenance solutions

#### C. Study Sample

In this study, five participants have been chosen to be interviewed. The sample is made up of three wind farm engineers and two operation and maintenance managers with working experience in the wind energy industry. These individuals were selected because they have practical and hands-on engagement with the Wind Turbines for their day-to-day functioning and maintenance. This affords them the much-needed, real-world knowledge regarding predictive maintenance systems, and the practicality of fine-tuning O&M efficiency, which fits well with research purpose and objectives. Instead, a smaller sample size was adopted to ensure that the interviews conducted were comprehensive and in-depth as well as to keep the themes manageable for analysis.

#### D. Data Analysis

The data obtained through the interviews was analyzed using thematic analysis as the approach. This approach is ideal for pattern finding and analysis or reporting themes within qualitative information. The use of thematic analysis in this particular study is beneficial because it enables the researcher to identify specific themes regarding the applicability of the predictive maintenance model and the fine-tuning of intelligent O&M systems. The steps involved in thematic analysis for this study were as follows:

TABLE I. STEPS OF THEMATIC ANALYSIS (SOURCE: AUTHOR)

| Step | Description |
| --- | --- |
| 1. Familiarization | Transcribing the interviews and reading the data to get an overall sense of the content. |
| 2. Coding | Identifying key patterns and labelling significant sections of the data with codes. |
| 3. Generating Themes | Grouping codes into broader themes that align with the research objectives. |
| 4. Reviewing Themes | Ensuring that the themes accurately represent the data by reviewing and refining them. |
| 5. Defining & Naming Themes | Clearly defining each theme and giving it a representative name. |
| 6. Reporting | Synthesizing and presenting the final themes in the context of the study's objectives. |

The approach of thematic analysis was deemed suitable, as it is a versatile method that has a high level of rigour to analyse qualitative data effectively to look at the experiences of the participants. This method is also useful for the detection of additional information, which is especially important in cumulative analysis of the PM and O&M optimization in wind power systems.

### IV. RESULTS AND ANALYSIS

Themes showed in the fig. 3 are taken from the interview guides that are going to be analyzed in this section.

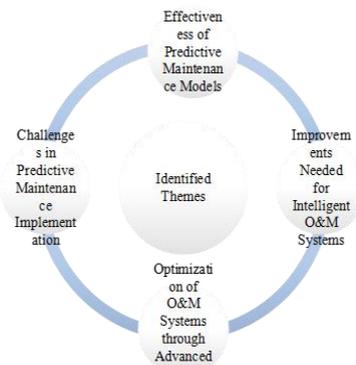

Fig. 3. Identified themes (Source: Author)

*A. Effectiveness of Predictive Maintenance Models*

Predictive maintenance models are essential in wind power systems to reduce and avoid system downtime due to failures. Many participants mentioned stages in which the majority of issues are identified, but the models were criticized for failing to detect minor gradual errors. Both the models also lack real time accuracy and compatibility with the older systems of turbines reducing efficiency. Some of the participant responses concerning the effectiveness of predictive maintenance models are provided below:

Farm Engineer

"Current models are helpful but require enhancements in the ability to identify minor problems before they turn into more significant ones. Although, with the application of predictive maintenance models, equipment downtime have been minimized prediction of small faults, which may lead to more catastrophic faults, is still a challenge."

Maintenance Manager

"The models work quite well, especially identifying large problems but they are less accurate especially during inclement weather. Real-time could be optimised to enhance the accuracy in avoiding unnecessary shutdowns."

Farm Engineer

"Other models have seen a decrease in downtime by up to about 20% thereby enhancing the efficiency of the turbines. Nevertheless, the false alarms occur more often and result in improper interventions and inefficient usage of resources that harm organizational productivity."

Maintenance Manager

"For large faults, the models are effective in minimizing downtime, but the models do not always capture small growing faults that may not be detected until it is too late. These issues can still lead to various operational problems."

Farm Engineer

"The models are relatively useful, especially for machines that are relatively young. Nonetheless, as for the older models, the level of accuracy of the predictions fall considerably combined with higher frequency of intervention and more unpredictable breakdowns."

*1) Analysis:* From the interviews, there was a clear indication of pros and cons of predictive maintenance models. A farm engineer noted that although the models are helpful, we need higher precision for spotting minor problems before they become severe. Another farm engineer said that the models fail to capture slow, progressive failures that affect turbines' effectiveness. A maintenance manager commented that while these models cut downtime by 20%, 'false alarms' constantly emerge, thereby resulting in unneeded actions. In general, participants recognized the usefulness of the models but pointed to the need for more accurate fault identification and less false alarms to ensure advanced effectiveness. Yang et al. [9] have also made similar observations, where predictive models have high success rates in cases of larger and more critical turbine faults but are less efficient at identifying relatively-small early-stage problems, especially in the older turbines. Likewise, Kusiak et al. [10] noted that false alarms, which arise due to sensor errors or changes in environmental factors, continue to be a chief challenge in the application of predictive maintenance. It is evident from these studies that predictive maintenance has come a long way in enhancing wind turbine performance, but there is still room for improvement in the models' accuracy in order to eliminate unnecessary iterations of maintenance.

*B. Challenges in Predictive Maintenance Implementation*

The implementation of the models of predictive maintenance is not easy to achieve because of certain barriers such as unreliable data, problems with the sensors, and false alarms. Some of the challenges highlighted by participants include the problem of overlaying new predictive technologies on older turbines, which has the effect of reducing their efficiency in preventing order and duration. Some of the participant responses concerning the effectiveness of predictive maintenance models are provided below:

Farm Engineer

"The main problem is the ability to merge new predictive models with old ones which is installed in the turbines, thus getting equivocal data quality. This compromises its effectiveness in ensuring that failures are well predicted in the entire fleet of turbines."

Maintenance Manager

"One of the major concerns would be the reliability of the sensors. Most of the time, wrong sensor readings are produced and as with any moving part, they affect the overall effectiveness of the model, which may lead to potential incorrect fault identification or missing a planned maintenance opportunity."

Farm Engineer

"It is for this reason that false positives are very common, leading to unnecessary maintenance actions. This disrupts the overall operational flow and puts resources into unnecessary tasks, which then adds to the maintenance costs."

Maintenance Manager

"Data overload and abundance of information where it becomes difficult to determine what information is relevant. One disadvantage is that the level of processing and generating information from a number of turbines becomes low, thus decision-making is slowed throughout the maintenance activities."

Farm Engineer

*"Inconsistency of the SCADA data especially that received from the remote areas is a big challenge to the model. This makes it difficult to identify and fix emerging faults, which can lead to system downtime."*

*1) Analysis:* The interviews showed several issues with the application of predictive maintenance. One of the farm engineers mentioned that while upgrading tended to incorporate newer but the predictive models that are produced are only compatible with recent turbines this has been drawback since it has lead to inconsistency in data quality. A farm engineer noted that too many false positives meant that machinery had to be repaired even though it was not faulty. In terms of maintenance, a maintenance manager stated that many sensors were not reliable due to the fact that they often produced wrong data. A maintenance manager highlighted a problem of information overload to the extent that it hampers decision making processes. Such issues include variation in SCADA data from remote areas and, thus, require the integration of the model, refined sensors, and better data management for more efficient predictive maintenance. These results are in line with the outcomes presented by Tian et al. [11], whose analysis revealed data inconsistency and integration difficulties when implementing predictive maintenance models for aged wind turbines. Similarly, Reder et al. [12] stated that, difficulties in sensor data interpretation and accuracy continue to prevail, resulting into complicated alarm management and additional operations. Consequently, there are significant opportunities to enhance existing solutions with adequate information processing and more advanced sensors to increase the accuracy of the predictive maintenance models in large wind farms.

*C. Optimization of O&M Systems through Advanced Technologies*

New technologies including digital twins, SCADA systems, and condition monitoring to add the level of optimization of O&M systems. Some of the issues highlighted by participants include increased energy output and better maintenance practices, but participants also noted that more research is needed to optimize how these technologies can be adopted in turbine production and operation.

Farm Engineer

*"The use of digital twin technology has been quite impactful. Inasmuch as it is set up to model real time operation of a turbine, it assists in improving efficiency and preventing failure through boosting maintenance and increasing energy generation."*

Maintenance Manager

*"The implementation of the modern SCADA systems with real-time data is one of the ways that has enhanced effective scheduling for maintenance. This technology ensures close tracking of the turbine operations; hence any inefficiencies can be noticed early for rectification leading to improved energy generation and less time out of service."*

Farm Engineer

*"Vibration analysis, combined with machine learning, has been particularly effective in identifying potential bearing failures. This predictive method helps extend the lifespan of turbine components and minimizes unexpected downtime."*

Maintenance Manager

*"Mild health indicator systems have gone a long way to help in early fault detection especially if there is a problem with the gearbox. Such systems also provide opportunities for more focused maintenance activity, excluding unneeded inspections, and improving turbine reliability."*

Farm Engineer

*"Technological advancements in this area have therefore made it possible for companies to adopt remote monitoring systems as a way of eliminating physical monitoring. They facilitate constant monitoring of turbine performance to enhance energy productivity while conducting maintenance, even in areas with poor accessibility."*

*1) Analysis:* The interviews revealed various innovative technologies, which have enhanced wind power generation turbines in several ways. A farm engineer explained that digital twin has helped in enhancing the performance of the turbines through maintaining the real time failures, hence improving the maintenance of the farms and energy produced. One of the farm engineers I interviewed said that using vibration analysis with machine learning could efficiently diagnose bearing failures thus minimizing the likelihood of downtimes. Maintenance managers positively acknowledged that SCADA systems have enhanced the timeliness of maintenance schedules and condition monitoring systems for fault description. Moreover, one of the farm engineers stated that the remote monitoring systems that have been adopted have reduced the need for physical inspections and have enhanced the overall energy management irrespective of the geographical area. These benefits identified and highlighted by the respondents are all captured by the current research. Shen et al. [13] confirmed that the utilisation of digital twin technology optimises turbomachinery maintenance through real-time fault prognosis and, therefore, minimal downtime. Likewise, Pandit et al. [14] underscored the necessity of SCADA systems and condition monitoring for maintaining descriptive schedules of equipment and identifying the fault in the budding stage, thereby enhancing the efficiency of turbines. All these technologies advanced operational performance according to the literature and wind farm operations as depicted by the participants.

*D. Improvements Needed for Intelligent O&M Systems*

Some of the key improvements needed as highlighted by the participants included; better AI models to ensure fewer false alarms, better sensors for real-time fault identification, and integration issues with older turbine systems in enhancing the efficiency, reliability, and predictive accuracy of the intelligent O&M systems.

Farm Engineer

*"Using real-time weather data might help increase fault detection accuracy in weather-dependent environments when*

*incorporated into the models. This would aid in formulating the best strategies to use depending on the environmental conditions in the operation of the turbines, hence, improving the efficiency."*

Maintenance Manager

*"This is because false positive rates are high and AI models require further calibration. This would reduce the number of times that the turbines were subjected to avoidable maintenance activities and expenditures thus increasing the efficiency of the needed resources with greater up time for the turbines."*

Farm Engineer

*"There is a need for the improvement of the technology for analyzing large amounts of data in a faster way. Enhanced speed of processing might reduce the time taken to identify faults thus reducing on the time taken by the turbine thus increasing its efficiency."*

Maintenance Manager

*"To enhance real-time fault detection, more advance sensors are needed. Higher accuracy from the sensors would mean that the O&M systems could detect and anticipate failures, thus cutting on instances of downtime and increasing efficiency."*

Farm Engineer

*"Previous designs of turbines require software solutions for improved compatibility with current O&M systems. It will help enhance forecasting reliability on a diverse range of vehicles, making performance uniform and minimizing the need for follow-up adjustments."*

*1) Analysis:* The interviews shed light on various areas where improvements are needed in intelligent O&M systems. One farm engineer suggested integrating real-time weather data into predictive models to improve fault detection accuracy under changing environmental conditions. Another farm engineer emphasized the need for enhanced data analytics tools to handle larger datasets efficiently. Maintenance managers pointed out the importance of fine-tuning AI models to reduce false positives and improving sensor accuracy for real-time fault detection. Additionally, a farm engineer stressed the need for software updates in older turbines to better integrate with modern O&M systems, enhancing overall performance. These suggestions are supported by recent research. A study by Harrou et al. [15] found that incorporating real-time weather data into predictive models significantly improves fault detection accuracy in varying conditions. Additionally, research by Daneshi et al. [16] highlighted the importance of advanced data analytics and enhanced sensor technologies in improving real-time fault detection, reducing false positives, and increasing overall operational efficiency. These improvements are essential for the seamless integration of modern O&M systems with older turbine models, ensuring performance that is more consistent and minimizing downtime, as mentioned by the respondents.

V. CONCLUSION

Wind energy is becoming more critical as society moves from the traditional technique of using fossil fuels as the major source of energy in electricity generation, but this has been an issue in the efficiency of the turbine. It is noted that predictive maintenance models and O&M systems equipped with cognitive functions are currently pointed as the adequate tools that should be implemented to minimize downtime, optimize annual energy production and guarantee the dependability of wind turbines. Still, the performance of these models and systems in practice is questionable, problematic issues like data quality and quantification of false alarms, as well as integration into systems with old types of turbines.

These challenges are the primary reasons for conducting this research to evaluate the predictive maintenance models and to identify the methods of improving optimization of intelligent O&M systems. Therefore, the current study utilized qualitative research approach through conducting structured interviews to five wind farm engineers as well as maintenance managers. Some of these participants, who have years of industry experience, offered rich descriptions of the application of these technologies in practice. The interviews conducted were analysed using thematic analysis to enable the study to identify pertinent themes about the challenges as well as possibilities of enhancing predictive maintenance and O&M systems

The results pointed out on the observation that major fault modelling for predictive maintenance are efficient in their senate and capability of a reducing downtime due to overall major faults while they fails to factor the minor gradual faults which brings inefficiencies to the overall operation. Further, issues like, malfunction of sensors, existence of false alarms, and the complexity of retrofitting new models with older turbines were often raised. Nonetheless, different technologies such as simulations of digital twin, systems for the control and data acquisition SCADA, and condition monitoring have greatly improved the schedule for maintenance and fault identification. Thus, these technologies show significant opportunities, however, certain developments are still required, especially, the enhancements of the real-time data integration, AI model, and the sensors. Eliminating these gaps will enhance the efficiency of the turbines, decrease general costs of operation, and improve the effectiveness of wind energy, which in return will enhance the utilization of renewable energy.

*A. Study Limitations*

The present research involved a focus group analysis of five people; this study does not give a comprehensive look at the staff experience of the wind energy business. Moreover, since this study falls into the qualitative category of research, it may present problems of external validity. Further research could include larger and more diverse samples and more quantitative approaches in order to better generalize knowledge of predictive maintenance and O&M.